\documentclass[10pt,a4paper]{article}
\usepackage{t1enc}
\usepackage[latin1]{inputenc}
\usepackage[english]{babel}
\usepackage{harvard}
\newcommand{\C}{\mbox{\rm\hspace*{0.6ex}\rule[0.2ex]{0.08ex}{1.2ex}\hspace{-0.7ex}C}}
\newcommand{\R}{\mbox{\rm I\hspace{-0.33ex}R}}

\newcommand{\M}{\mbox{\rm I\hspace{-0.31ex}M}}
\newcommand{\h}{\hbar}
\newcommand{\1}{\mbox{\rm 1\hspace{-0.6ex}I}}
\newcommand{\beq}{\begin{equation}}
\newcommand{\eeq}{\end{equation}}

\begin{document}
\begin{center}
\subsection*{Local spinor structures  in  
 V. Fock's  and H. Weyl's work  on the Dirac equation (1929)  }
{\em Erhard Scholz, Wuppertal}
\end{center}
\begin{abstract}
In early 1929, V. Fock (initially in collaboration with D. Iwanenko) and H. Weyl developed independently from each other a general relativistic generalization of the Dirac equation.  In the core, they arrived at the same theory by the introduction of a local (topologically trivial) spinor structures and a lifting of the Levi-Civita connection of underlying space-time. They both observed, in slightly different settings,  a characteristic underdetermination of the spin connection  by a complex phase factor, which gave the symbolical possibility for a reformulation of Weyl's old (1918) idea to characterize the electromagnetic potential by a differential form transforming as a gauge field. Weyl and Fock realized the common mathematical core of their respective approaches in summer 1929, but insisted on differences in perspective. An interesting difference was discussed by Weyl in his Rouse Ball lecture in 1930,. He contrasted the new type of unification strongly to the earlier geometrically unified field theories (including his own). He was quite explicit that he now considered his earlier ideas on geometrization of ``all of physics'' as premature  and declared that the new, more empirically based approach would have to go a long way before it could be considered as a true "geometrization" of matter structures.

\end{abstract}

\subsubsection*{Introduction}
In the early 20th century the most important impact of mathematical physics on geometry came from relativity theory. Historical and philosophical questions of this interplay have been discussed at various occasions.\footnote{Among them \cite{Boi/Flament} \cite{Gray:Universe}.}
The rise of quantum physics brought about a second shift, philosophically, technically  and conceptually much deeper, for the relationship of geometry to physics. It started in the late 1920s,  gained momentum in the second half of  the past century and began to dominate the  image of  knowledge for the deeper levels of  physical geometry  during its last two  decades.\footnote{For a first historical exploration see \cite[section V]{Cao:QFT}, in particular J. Stachel's introductory remarks.} Other contributions to these conference proceedings are evidence  for the actuality  of this recent  and ongoing shift in our understanding of physical geometry,  which  is far from completed and continues to be an open-ended and controversial project.\footnote{Cf. contributions  of M. Atiyah and  A. Connes to this volume. }

An important turn in the relationship between relativity, quantum mechanics and field theory, which also  sheds  light on  the nature and role of geometry in this conceptual complex, was  initiated  by Hermann Weyl and Vladimir Fock  in early  1929. They   both started to investigate  (generalized) Dirac  fields  in the context of general relativity by the introduction  of local spinor structures on Lorentz manifolds.  This topic was  taken up anew in the 1960s from a global point of view.\footnote{Tthe role of the Dirac operator for the interplay between differential geometry and topology in the last third of the century is being discussed in J.-P. Bourgignon's contribution to this volume.}

Up to the end of the 1920s mathematical physicists had  essentially two symbolic tools for the represention of physical fields at their disposal: vectors/tensors (including differential forms) and linear connections (mostly but not always affine), most important among them, of course, the Levi-Civita connection of general relativity (GRT).  After 1918  H. Weyl  tried to convince physicists and mathematicians for some time to use another type of connection ({\em length connection}) in combination with a conformal (class of) Lorentz metric in his first, strictly metrical gauge geometry.\footnote{This approach is discussed,  from a more recent point of view, by   P. Cartier's  in his contribution to this volume.} 
Most physicists who considered  Weyl's length connection at all referred to it as just another differential 
1-form $\varphi = \sum \varphi _i dx^i$ with a peculiar, perhaps even strange, transformation behaviour. 

 In the early 1920s A. S. Eddington started to build his attempts towards a unified field theory  of electromagnetism, gravitation and matter using general  affine connections (not necessarily derived from a metric); and Einstein joined him for a while from 1923 onward.  
These activities were part of a broader move towards unified field theories (UFT's) with  a first high tide in the 20s of the last century, which has been studied historically, among others, by Vladimir Vizigin \cite{Vizgin:UFT} and, more recently and in a different  methodological approach, by Catherine Goldstein and Jim Ritter \cite{Goldstein/Ritter:UFT}.\footnote{Another high tide, in a different historical/scientific context and with changed  conceptual/symbolical approaches, started in the 1970s. It has not yet found the detailed and critical historical investigation it deserves, although  work has started \cite{Cao:FTs}, \cite{Morrison:Symmetries}, \cite{Galison:Strings}, \cite{Oraif/Straumann}. }
V. Vizgin presents the relationship of  UFT and quantum physics (QP) as one of competing research programs  mutually influencing  each other. The introduction of local spinor structures by Fock and Weyl in 1929 is a beautiful example for his case.
 Both, Weyl and Fock, were struck by the early successes of the Dirac  equation for the explanation of the motion of the electron   and attempted an integration of GRT and the Dirac field. In such an attempt they were not alone. Other authors, like Wiener and Vallarta,   attempted a similar integration along different lines, building upon Einstein's recent theory of ``distant parallelism''. They attempted to adapt the Dirac field to a framework of classical UFT's  that soon turned out to be too restrictive.

Weyl and Fock, the latter  after an initial phase of sympathizing with distant parallelism, pursued an approach  of a  covariant differentation of spinor fields derived from the underlying Levi-Civita connection,  in contrast to the distant parallelism program. Both realized that, in doing so, an underdetermination of the ensuing spinor connection  led  naturally  to an  additional $U(1)$-symmetry. They used the latter for a representation of the electromagnetic field comparable to, although slightly different from, Weyl's earlier approach using a  length connection. Thus they arrived  at a geometric-analytical structure in which the actual knowledge of gravitation, electromagnetism and the basics of the quantum  theory of the moving electron could be represented in an integrated form.\footnote{For a discussion of Weyl's 1929 work on gravitation and the electron see also \cite{Straumann:DMV}.} Both authors posed the  question  how geometry might be brought into agreement with quantum physical knowledge of their time. They  arrived at strongly diverging evaluations  as to what they had achieved in this respect and what geometrization of quantum physics might  mean at all   (last section). 

Before I discuss Weyl's and Fock's respective approaches and differences with respect to ``quantum geometry'', I want to sketch the  background of common knowledge from which they started and outline their 1929  work.

\subsubsection*{Setting the stage in the later 1920's for Weyl and Fock}
During the 1920s  the constitutive conditions for the mathematization of geometry and matter   changed deeply. In the middle of the decade (1925/26)  the  ``new'' quantum mechanics took shape, with its different versions,  in central aspects compatible, although at least  historically and conceptually not completely equivalent,  put forward by Heisenberg/Born/Pauli,  Schr\"odinger and Dirac.\footnote{For  a general picture see \cite{Rechenberg:QM},  \cite{Pais:Inward} and  \cite{Hendry:Bohr-Pauli}.} Continuing this turn in late 1926, W. Heisenberg started to investigate the symmetry of atomic electrons  using surprisingly  old-fashioned mathematics, Serret's {\em Alg\`{e}bre sup\'erieure} from 1879.  But already in the following year the two young Hungarians,  E. Wigner and J. von Neumann, working in Berlin and G\"ottingen, applied group representation methods for this goal, as did H. Weyl  in a lecture course devoted   to this subject in the winter semester 1927/28 at the ETH Z\"urich.\footnote{\cite[488ff.]{Mehra/Rechenberg:Completion_I}.}

Still in 1926, W. Pauli attempted to characterize the new hypothetical electron   ``spin'' in terms of quantum mechanical symbolism and introduced  a pair of ``wave'' functions
 ($\psi _1 (x), \psi _2 (x)$), $x \in \R^3$, and Hermitian matrices, which later were given his name,
\[ \sigma_1 =  \left(  \begin{array}{cc} & 1\\ 1& \\ \end{array}  \right) , \; \;
\sigma_2 =  \left(  \begin{array}{cc} & -i \\ i& \\ \end{array}  \right) , \; \;
\sigma_3 =  \left(  \begin{array}{cc} 1& \\ &-1 \\ \end{array}  \right) . \] 
Pauli proposed  to  represent the electron spin by the three component operator
\[ \sigma = \frac{1}{2} \h (\sigma_1, \sigma_2, \sigma_3) , \; \; \h = \frac{h}{2 \pi} \, . \]

Like Heisenberg, Pauli did not think in terms of group representations at that time; he constructed his two-valued wave functions from the Klein-Sommerfeld theory of the spinning top and the complex representation of rotations by Cayley-angles. 
That was an ingenious  and  mathematically momentous  move towards what  little later turned into (Euclidean or relativistic) spinors, although Pauli's hopes to come to a direct   explanation of  the fine structure of    the hydrogen spectrum were not fullfilled at the time.\footnote{\cite[289ff.]{Pais:Inward}.} Even the first attempts  in 1926 and 1927  to take relativistic effects into account, spinless  (Klein-Gordon) or with spin (Darwin), were  no more successful  in this respect.\footnote{\cite[44ff.]{Kragh:AHESDirac}, \cite[280ff.]{Mehra/Rechenberg:Completion_I}.} 
The situation changed completely  in January and February 1928 when  Dirac proposed to use  4-component complex-valued ``wave'' functions $\psi (x) = (\psi_1(x), \psi_2(x),\psi_3 (x), \psi_4 (x))$ ($x$ in Min\-kow\-ski-space $\M$) in two successive publications \footnote{\cite{Dirac:1928_I/II}.}. The  $\psi$- function had to obey the ({\em Dirac})   equation
\beq i \h \sum_{\alpha = 0}^3  \gamma ^{\alpha} \frac{\partial}{\partial x^{\alpha}} \psi = 
m_0 c \psi \eeq 
 with (Dirac) matrices   $\gamma ^{\mu}$  satisfying the   relations 
$ \gamma ^j  \gamma ^k +  \gamma ^k  \gamma ^j = \delta _{j k}$ and expressible, e.g., in the form
\[ \gamma ^0 =  \left(  \begin{array}{cc} \1 &\\ & -\1  \end{array}  \right) ,  \; \; 
    \gamma ^j =  \left(  \begin{array}{cc}  & \sigma_j \\  -\sigma _j &   \end{array}  \right) ,  \; \;
   1 \leq j \leq 3 ,  \]
with ($2 \times 2)$-unity matrix $\1 $ and Pauli matrices $\sigma_j$.\footnote{Dirac used a slightly different presentation of the matrices than the one given in the text. For a detailed investigation of Dirac's work see \cite{Kragh:AHESDirac} or \cite{Kragh:Dirac}.} 

Thus things looked quite different for Weyl in the late 1920s   from what they had been at the end of  his first phase of activity in mathematical physics early in the  decade. Already in late 1920 he had lost confidence in theories of matter  by unification of classical fields according to the  Hilbert/Mie approach, including   his own one built upon the length gauge.\footnote{See  \cite[chap. V]{Skuli:Diss} or \cite{Scholz:Infgeo}.}  While expecting new insights from the rising quantum mechanics, he concentrated on more conceptual or purely mathematical research fields: the analysis of the space problem about 1922/23 and representation theory of Lie groups during the years 1924 to 1926.\footnote{\cite[Part IV]{Hawkins:2000}.}
Weyl kept  well informed on the  ongoing development during the crucial years for quantum mechanics in the middle of the decade, drawing upon his close scientific relationship with Pauli  (1924 -- 1928 at Hamburg university), dating from their cooperation on unified geometrical field theories  in  the early 1920s. Moreover he had   contacts with E. Schr\"odinger who taught at the university in Z\"urich between  1921 and 1927. He appararently felt challenged  to contribute to the  conceptual and mathematical clarification of the framework of the ``new'' quantum mechanics, in particular from the point of view  of  {\em unitary geometry} (Weyl's title for the first part of his lecture in 1927/28) and the use of  {\em representation theory}   of (Euclidean) rotations and permutation for atomic line spectra, Pauli's non-relativistic spin, and mechanism of  molecular binding forces.  

In winter 1927/28 Weyl had a chance to take up the challenge. Both theoretical physicists working at Z\"urich had accepted outside calls and had left: P. Debye changed from the ETH to the university Leipzig and E. Schr\"odinger from the local university to Berlin. Weyl decided to change the subject of  a lecture course initially planned and announced on (pure) group theory to one on {\em Gruppentheorie und Quantenmechanik (Theory of Groups and Quantum Mechanics}. Notes were taken by his assistant F. Bohnenblust and published, after revision and extension, in August 1928 as a book  \cite{Weyl:GQM}, which  in the sequel will be abbreviated as {\em GQM}. In this second book on mathematical physics, Weyl was more cautious than he was  in {\em Raum - Zeit - Materie} \cite{Weyl:RZM} in his expectations of how his contributions might be received by the workers in  the  field. In the preface to the new book, he remarked:
\begin{quote}
It is the second time that I dare to turn up with a book   which belongs only partly to my own speciality, mathematics, and partly to physics. \ldots I just cannot avoid to play the role of a messenger (often undesired, as I have experienced sufficiently clearly) in this drama of mathematics and physics -  fertilizing each other in the dark, although from face to face  preferring not to recognize and even renouncing each other. \cite[Vf., my translation, E.S.]{Weyl:GQM}\footnote{Not translated in the English edition by H.P.Robertson.}
\end{quote}
Weyl was not alone in this "role of a messenger" as he realized during the preparation of the lecture notes for  publication. Other authors  started  in 1927 and 1928 to use group representations in quantum mechanics, among them, most importantly from the mathematical point of view,  J. von Neumann and E. Wigner. Also on the physical side, things  changed  rapidly.  Dirac published his papers on the relativistic theory of the electron at the end of the winter semester, in January and February 1928. The impact was enormous and  were sufficient reason  for Weyl to  add to his book  a whole new  passage on Dirac's equation 
\cite[1st ed., \S \S 39--41]{Weyl:GQM}.

Another remark in his lectures of 1927/28 leads directly to our geometrical topic.\footnote{This passage  was published only in the first edition of \cite{Weyl:GQM}, no longer in the second edition of 1931 and the English translation.}
 Weyl's gauge idea from 1918, originally linked to a length calibration and  `` infinitesimal length transport'' characterized by a 1-form   $\varphi = \sum \varphi _i dx^i$ was rephrased in a quantum mechanical setting by E. Schr\"odinger, still  in a length calibration interpretation \cite{Schrödinger:1923}, and after the rise of the ``new'' quantum mechanics by V. Fock and F. London  in the context of Kaluza-Klein theory of  quantum mechanics \cite{Fock:1926,London:1927}. The core of their respective arguments dealt with  ``gauging''  a wave function $\psi (x)$ by a point-dependent  phase factor $ e^{i \lambda (x)}$ (with $\lambda \in \R$)   to
 $\tilde{\psi }(x) = e^{i \lambda (x)} \psi (x)$.  The differential of the purely imaginary phase factor, used in Weyl's 1918 theory to ``gauge-transform'' length connections, could now be used  to transform electromagnetic potentials $\varphi_j$ a little  more convincingly

Weyl endorsed this recontextualization of his original gauge idea when he discussed the Schr\"odinger equation in 1927/28. Probably he had read only the papers by Schrödinger and London,  which he cited, not Fock's; but London  was aware of and built upon \cite{Fock:1926}.\footnote{ \cite[293]{Vizgin:UFT}.} 
He remarked that the Schr\"odinger equation
\beq  i \h \frac{\partial \psi}{\partial t} = H \psi \, ,\eeq
containing the Hamilton operator
\beq H = \frac{1}{2m} \sum p_j^2 + V(x) \eeq
with potential $V$ and momentum operator $p_j = \frac{\h}{i} \frac{\partial}{\partial x^j}$ for a chargeless particle, is adequately modified  by using the covariant derivative $\partial_{\varphi}$ with respect to  a potential connection $\varphi = (\varphi_j)$, if a charged particle in field of potential $\varphi$ is considered. Then  the momentum operator 
becomes 
\beq p_j = \frac{\h}{i} \left( \frac{\partial}{\partial x^j} + \frac{i e}{\h} \varphi_j \right) , \; \; \;  \; i = \sqrt{-1} \; ,  \eeq
and the Hamiltonian of the Schr\"odinger theory  for the motion of a particle of charge $e$ in an electromagnetic field of potential $\varphi$ results. Weyl observed that now:
\begin{quote}
The field laws satisfied by the potentials $\psi $ and $\varphi$ of the material and the electromagnetic waves are invariant under simultaneous substitution of 
\[ \psi \; \mbox{by} \; e^{i \lambda} \psi , \; \; \; \varphi_{\alpha} \; \mbox{by} \; 
 \varphi_{\alpha} - \frac{\h}{e}\frac{\partial \lambda }{\partial x_{\alpha }} \]
\ldots \cite[1st ed. 87f.]{Weyl:GQM}
\end{quote}
He commented that this ``principle of gauge invariance'' was quite analogous to the one he had  postulated in 1918 ``by speculative reasons to gain a unified theory of gravitation and electromagnetism'' and continued:
\begin{quote}
\ldots But now I believe that the gauge invariance does not couple electricity and gravitation, but rather {\em electricity and matter} in the mode presented here. How gravitation according to the general theory of relativity can be included is still uncertain. \cite[1st ed. 88]{Weyl:GQM}
\end{quote}
Thus Weyl   proposed  more than a technical adaptation of his old gauge idea  to the new framework of QP.  In classical UFT the goal was to unify force fields as such in a coherently geometrized, often highly speculative, ``a priori'' manner,   and to derive matter structures from them; here Weyl indicated  a new paradigm centering around the search for conceptual and mathematical structures which link forces to matter fields, without reduction of one to the other and with strong input from experimental evidence. 

Classical UFT was, of course, still quite alive at that time. In 1928 A. Einstein  turned towards  ``distant parallelism'' for his latest approach to unification. He assumed or postulated, that, in addition to the Levi-Civita connection of the Lorentz metric,  an integrable, curvature free, orthogonal connection $\Delta ^i{}_{jk}$ with torsion ($\Delta ^i{}_{jk} = - \Delta ^i{}_{kj}$) is given, which he usually described by a globally parallel system of orthogonal frames. With respect to such an additional structure it was meaningful to consider constant, i.e. point independent, rotations. Although Einstein did not intend so, his additional structure allowed  a formulation of the Dirac equation in the framework of GRT with distant parallelism and stimulated other physicists to do so. 

V. Fock and his Leningrad colleague D. Ivanenko started to explore such an approach in a joint paper submitted   to {\em Zeitschrift f\"ur Physik} in March 1929.\footnote{March 25, 1929.}
 They hoped to find  some  ``bridge'' between gravitation and quantum theory.\footnote{For the group of young relativists in Leningrad see \cite{Gorelik/Vizgin}, for the early involvement in QP \cite{Frenkel/Gorelik:Bronstein}. More on Fock in \cite{Gorelik:Fock}.}  They started with a formal construct of a  linear expression in the Dirac matrices, $ds = \sum _j \gamma_j dx^j$, which they tried to interpret  as a matrix valued metric form of some new  ``linear quantum geometry''. From that point of view they  hoped  to find a kinship  between Einstein's field of distant parallelism and the new ``linear geometry''   \cite[801]{Fock/Ivanenko:ZPh}. During the following months Ivanenko and Fock realized that  the linear structure of the new geometry  could better  be understood as a covariant derivative  of the 4-component complex wave functions which they called ``semi-vectors'', the later spinors.\footnote{The terminology of  ``semi-vectors'' was proposed by L. Landau.} Still they  called the  geometry they were heading for  ``g\'eom\'etrie quantique lin\'eaire''   \cite{Fock/Ivanenko:CR,Fock:CR1929}.\footnote{\cite{Fock/Ivanenko:CR} was submitted May 22, 1929.} V. Fock continued to explore the terrain  and realized soon that the new covariant derivation of spinors had a much closer kinship with a Weylian phase gauge than with Einstein's distant parallelism. He presented his findings in two articles (no longer co-authored by Ivanenko) to {\em Physikalische Zeitschrift} and {\em Comptes Rendus} 
 \cite{Fock:CR1929,Fock:ZPh1929}.\footnote{\cite{Fock:CR1929} dated June 24, \cite{Fock:ZPh1929} July 5, 1929.} He thus arrived at a theory combining gravitation, Dirac field, and electromagnetism, which overlapped in large parts with what Weyl achieved in early 1929 when he continued research along the  lines indicated in {\em GQM}.

\subsubsection*{Weyl's and Fock's  local spinor structure}
 Weyl  left Z\"urich in September 1928 for Bologna (ICM) and Princeton where he spent a year as  reseach professor in mathematical physics.\footnote{\cite[107ff.]{Frei/Stammbach:Weyl}.} There he could continue, among other things, his research on the Dirac equation  in general relativity. The approach of distant parallelism did not  appear at all convincing to him. He   considered it to be a completely ``artificial'' device and  looked for a combined structure of  GR and the Dirac equation from the point of view of ``purely infinitesimal'' geometry, which now had  to be refined and extended in the light of new physical knowledge. In February 1929 Weyl submitted a first sketch of methods and   results under the title  {\em Gravitation and the electron}  to the {\em Proceedings of the National Academy of Sciences} \cite{Weyl:NAS1929}. Three months later he delivered a more extended exposition to {\em Physikalische Zeitschrift}  \cite{Weyl:ZPh1929}.\footnote{Submitted, May 8, 1929.} At that time he could not know of Fock's parallel work, nor did he know of it  when he wrote his third paper on the topic in early summer \cite{Weyl:Rice1929}.

Fock, on the  other hand, got to  know of Weyl's new researches \cite{Weyl:NAS1929} only after he finished  his  own article for  {\em Physikalische Zeitschrift}. He accepted the common mathematical core of their respective approaches, but emphasized the  differences from the physical point of view in a postscript \cite[276f.]{Fock:ZPh1929}.
Weyl apparently got to know Fock's work in summer 1929 and was so fond of the common  features of their work that he considered it as establishing essentially one and the same theory. He thus referred to it  in the preface to the second edition of {\em GQM}  as the ``general relativistic formulation of the quantum laws, which have been developped by Mr. V. Fock and  the author [Weyl himself]'' \cite[vii, 2nd edition 1930]{Weyl:GQM}.\footnote{Weyl  saw no chance to give an exposition of this theory in the book {\em GQM}. In the second edition he rephrased, however, his discussion of the representation of the Lorentz group and of the special relativistic Dirac equation, in particular the decomposition of the 4-dimensional spinors into irreducible 2-dimensional representations.}

Fock and Weyl applied  the method of (pseudo-) orthogonal moving frames in Lorentzian 
space-time $M$, i.e. they supposed  an
 \[\mbox{orthonormal frame  of tangent vectors (ONF): \hspace{3mm}} e(\alpha,x) , \; \;
 0 \leq \alpha \leq 3 ,  \]
 in  each point $ P \in M$ with coordinates $ x =(x^0, \ldots, x^3)$ (depending differentiably on the point). 
 Tangent vectors $v$ at $x \in M$ could thus be represented in components referring  to the coordinate basis  $ (\xi^j )$, or in components  with respect to the ONF $(\xi (\alpha )$ in  Weyl's notation):
\beq v = \sum_{j=0}^3 \xi^j \frac{\partial}{\partial x^j} = \sum_{\alpha =0}^3 \xi (\alpha ) e(\alpha ,x) \, . \eeq
 Besides (differentiable) change of coordinates,  changes of the ONF from $e(\alpha ,x) $ to $e'(\beta ,x)$ ($0 \leq \alpha , \beta  \leq 3$) had also to be taken into account. The latter  were given by point-dependent 
Lorentz-rotations $\vartheta (x)$ represented by matrices (as the ONF's were  given in components with respect to a local coordinate system):
\[ \vartheta (x) =  (\vartheta^{\alpha }_{\beta }) \in SO(1,3) \,  . \]
The parallel transport of a frame by the Levi-Civita connection  $\Gamma^i{}_ {j k}$ could be expressed in terms of ``infinitesimal rotations'' $o$ depending linearly on infinitesimal displacements $dx =(dx^j )$ in space-time
\beq  o ^{\alpha }_{\beta }  = \sum_k  \omega ^{\alpha }_{\beta k}  dx^k \, . \eeq
 In more recent terminology: By means of the ONF's Fock and Weyl reduced the group of the affine connection  $\Gamma^i{}_ {j k}$ to the orthogonal group, and characterized parallel transport in $M$ by the resulting orthogonal connection $ \omega ^{\alpha }_{\beta k}$.

In the late 1920s  this was standard knowledge. The  idea of ONFs had already been introduced by Ricci and Levi-Civita  in 1900; it had been worked out by differential geometers in the 1920s, most prominent among them E. Cartan (in lectures from 1926/27  published as \cite{Cartan:Lecons_ONF}), J.A. Schouten,  R. Weitzenb\"ock, L.P. Eisenhart (in monographs 1926 and 1927). Moreover, orthonormal frames played a central role in Einstein's theory of ``distant parallelism'', from which  Fock (and Ivanenko) took the idea.\footnote{In his main article Fock referred, however, also to \cite{Eisenhart:1926}  \cite[263, footnote]{Fock:ZPh1929}.} Fock (still in his cooperation with Ivanenko) and Weyl realized that  reduction of the Levi-Civita connection to the orthogonal group by the ONF method allowed one to introduce covariant differentiation of spinors.\footnote{For simplicity, I will no longer always add in the sequel  Ivanenko to Fock, even in case that concepts appeared already in their joint work.} 
Weyl explained clearly that  the orthogonal reduction of the connection was necessary in  this context,  because ``Dirac's quantity'' $\psi$
\begin{quote}
\ldots corresponds to a representation of the orthogonal group which cannot be extended to the group of all linear transformations. The tensor calculus is consequently an unusuable instrument for considerations involving $\psi$. \cite[219]{Weyl:NAS1929}
\end{quote}
 For Weyl, this group-theoretic consideration was of great importance. In the early 1920s he had analyzed the role of tensors from the point of view of group representions and found out that all irreducible representations of $GL(n,\R)$ with a specified permutation symmetry can be characterized by tensors over $\R^n$.\footnote{See \cite[440ff.]{Hawkins:2000}.} In a language  closer to physicists he explained  more in detail:
\begin{quote}
Vectors and [tensors] are so constructed that the law which defines the transformation of their components from one Cartesian set of axes [ONF] to another can be extended to the most general linear transformation, to an affine set of axes. That is not the case for [the] quantity $\psi$, however; this kind of quantity belongs to a representation of the rotation group which cannot be extended to the affine group. \cite[234]{Weyl:NAS1929}
\end{quote}

He admitted that the ONF method used by him resembled Einstein's latest appproach in formal aspects, but insisted that this was only a superficial coincidence.
\begin{quote}
But here there is no talk of ``distant parallelism''; there is no indication that Nature has availed herself of such an artificial geometry. I am convinced that if there is a physical content in Einstein's latest formal development it must come to light in the present connection. 
\end{quote} 
And he added a reason that went  beyond purely mathematical considerations:
\begin{quote}
It seems to me that it is now hopeless to seek a unification of gravitation and electricity without taking material waves into account. \cite[219]{Weyl:NAS1929}
\end{quote}

Dirac had  shown that the {\em equation} of the free electron expressed in  $\psi$ is invariant under Lorentz transformations  without  asking for the underlying reprentation of the Lorentz group,\footnote{\cite[310ff.]{Dirac:1928_I/II},  discussed in \cite[57f.]{Kragh:AHESDirac}.} 
but other authors did so  immediately later.  F. M\"oglich calculated the complex $4\times 4$-matrices for the ``Dirac-quantity'' corresponding to a given Lorentz transformation     \cite{Moeglich:1928}, and J. von Neumann discussed the resulting relation
\begin{eqnarray*} \Lambda: \;  SO^+ (1,3)  & \longrightarrow & GL(4,\C)    \\
 o &  \longmapsto & \Lambda (o)
 \end{eqnarray*}
as a ``(multivalued!) 4-dimensional representation of the Lorentz group'' \cite[867]{Neumann:Dirac}.  Von Neumann emphasized, very much like Weyl,  that something essentially new was introduced  into mathematical physics:
\begin{quote}
The case of  a quantity of 4 components which is no 4-vector  has never occurred  in relativity theory, the Dirac $\psi$-vector is the first example of this kind. (ibid.)\footnote{Translation E.S.}
\end{quote}
Thus, immediately after Dirac's publications on the ``spinning'' electron, theoretically minded authors realized that the new ``Dirac quantity'' (Weyl), the ``$\psi$-vector''  (von Neumann), or the ``semi-vector'' (Fock, Landau e.a.) was more than just another technical device, but   led to a  {\em conceptual innovation}  for mathematical physics. Change of reference systems in special relativity (``Cartesian systems of axes'' as Weyl would say) by a Lorentz transformation had to be represented by $\Lambda (o)$ in the $\psi$-space in a way 
 that could not be extended to general linear transformations and thus  could not,  in a straight-forward manner,   be transferred to general relativity.

 At the time when Fock and Weyl approached the problem of a general relativistic formulation of the  Dirac equation, the young algebraist B.L. van der Waerden established an algebraic calculus for all possible quantities appearing in any representation of the Lorentz group. His contribution was meant as a sort of service to the physicists, stimulated by a question of  P. Ehrenfest who had posed  the question to design such an algebraic calculus. Van der Waerden picked up the terminology``spinor'' from Ehrenfest and gave him a broad audience  \cite[100]{Waerden:Spinoren}. In this work he built upon Weyl's exposition of the representation theory of the Lorentz group in {\em GQM}.

Distinct from other work  about 1929, Fock and Weyl admitted  point-dependent (Lorentz-) rotations of ONF in  space-time, $o(x) \in SO^+ (1,3)$, differentiably depending on $x$, inducing point-dependent transformations $\Lambda (o(x))$ of the spinor space. While Fock immediately headed for the covariant derivation of a spinor (``semi-vector''), Weyl made the underlying invariance idea explicit. He stated for  the ``laws'' that  would be characterized  by an action principle and by differential equations derived from it:
\begin{quote}
The laws shall remain invariant when the axes in the various points $P$ are subjected to arbitrary and independent rotations. \cite[219]{Weyl:NAS1929}
\end{quote}
Variational equations  were thus required  to be invariant under simultaneous transformations 
\begin{itemize}
\item[---] of vectors/tensors by Lorentz rotations $o(x)$
\item[---]  and of the spinors under $\Lambda (o(x))$.
\end{itemize}

In this way, Weyl and Fock  introduced and started to study a {\em  local spinor structure} on the underlying space-time manifold $M$. Both authors {\em used} local change of coordinates in the spinor space $\Lambda (o(x))$ (the change of trivialization in later language) accompanying a change of ONF's $o(x)$,  and Weyl  {\em discussed} its conceptual role quite clearly, although of course not yet applying the terminology of local bundles trivialization.

 Weyl did not mention, however,  that for a globalization of the procedure the topology of the $M$ might play a role. Such questions of global existence of an ONF (presupposing parallelizability of $M$), were  posed and answered  only in  the 1930s   by the young generation of  topologists (E. Stiefel, H. Whitney), apparently stimulated by Einstein's use of (local) ``distant parallelism'', not by  local spinor structures of Fock and Weyl.  Global  questions for spinor structures  were   taken up   still another generation later and became a research topic only  in the 1960s.\footnote{See P. Bourgignon's contribution, this volume.} Weyl, in his 1929 articles, did not even  indicate that there might be an open and challenging question in the relationship between spinor structures on $M$ and its topology.

Of immediate interest, for our authors, was the introduction of an ``infinitesimal displacement of semi-vectors'' (Fock) or the ``invariant change $\delta \psi$ on going from the point $P$ to a neighbouring point $P'$'' \cite[221]{Weyl:NAS1929}, i.e. in modern terminology the introduction of a connection and  parallel transport in a local spinor structure,  lifted from the Levi-Civita  connection in the underlying Lorentz manifold. On this point the two authors applied  slightly different approaches; Weyl's  approach was, as one may expect, more conceptual  and Fock's more calculational.

Considering two (infinitesimally) ``neighbouring'' points $P,\; P'$ with coordinates 
$x =(x^0 ,\ldots , x^3 )$  and $x' = ({x'}^{0} , \ldots , {x'}^{3})$ differing by an ``infinitesimal displacement'' $dx = (dx^0 , \ldots , dx^3)$ Weyl argued that parallel displacement of a  frame 
 $\{ e(\alpha ,P) \}$ from $P$ to $P'$ leads to an infinitesimally rotated  frame   $\{ e'(\alpha ,P') \}$  described by an infinitesimal rotation $o = \omega  (dx) $ with respect to  the ONF-system $\{ e(\alpha ,P') \}$  in $P'$, in slightly metaphorical notation
\beq \{ e'(\alpha ,P') \} - \{ e(\alpha ,P)\} = \omega  \cdot \{ e(\alpha ,P') \} \eeq
(compare equation (6) ). The  representation $\Lambda $ induces an infinitesimal tranformation $dE$ (Weyl's notation) in $gl (n,\C)$, which depends linearly on $dx$
\[  dE = \Lambda (o) = \Lambda \omega (dx)  \]
 The ``differential $\psi (P') - \psi (P)$'', i.e. $d \psi = \sum_j \frac{\partial \psi}{\partial x^j} dx^j$,  had  to be modified accordingly to give  the {\em covariant differential} $\delta \psi $ of  $\psi$ \cite[221]{Weyl:NAS1929} \cite[253f]{Weyl:ZPh1929}: 
\beq \delta \psi = d \psi  + dE \cdot \psi    \, .  \eeq
This conceptually clear description of the covariant differential, had the advantage that  in Weyl's discussion $\Lambda$  could stand for {\em any representation of the Lorentz group}, not just Dirac's original 4-dimensional one.

 Weyl realized of course, as did von Neumann in 1928, that Dirac's representation can be decomposed into two irreducible representations $\rho $ and $\rho ^+$ (which generate all finite dimensional representations of $SL(2,\C)$ by tensor products and direct sums). He gave a beautiful geometrical description of the 2-valued inverse of the covering map\footnote{\cite[247f.]{Weyl:ZPh1929}.} 
\[ SL(2,\C)  \longrightarrow SO^+(1,3) \] 
and took $\rho $ as the identical representation of $SL(2,\C)$  and
 $ \rho ^+ = \;  ^t{}\bar{\rho }$ its adjoint. 
Then he could write Dirac's representation (up to a permutation of $\psi$-coordinates) as 
\beq   \Lambda  \cong \rho  \oplus \rho ^+  ,\eeq 
and wrote the 4-spinors  (after a linear transformation) as 
$(\psi^+_1, \psi^+_2, \psi^-_1, \psi^-_2)  $.

Fock analyzed the condition (incorporated by Dirac into his new symbolic game) that the $\psi$-functions  get their physical meaning from the condition that the evaluation map
\[  \psi \longmapsto (a^0, \ldots, a^3) \; \; \mbox{with \hspace{2mm}} 
a^j = < \gamma ^j \psi , \psi > \; , \; \;  0 \leq j \leq 3, \]
leads to a vector $(a^j)$. Therefore it was natural to postulate that  ``changes of a semi-vector $\psi$ under an infinitesimal parallel displacement'' are  compatible with  parallel displacement of vectors. This  allowed him to compute    matrices  $C_l \in GL(4,\C)$  which describe such compatible ``infinitesimal changes of semi-vectors'' (the parallel displacement in the local spinor structure). In his own  representation $\tilde{\gamma }^j $ of the Dirac matrices Fock derived the condition
\beq  C_l = \frac{1}{4} \sum_{j,k,l}   \tilde{\gamma }_j \tilde{\gamma }^k
 \omega ^j{}_{k l} + i \phi _l  \; , \; \; \mbox{with \hspace{2mm}}
  \tilde{\gamma }_j = \sum_k \epsilon _{ j k} \tilde{\gamma }^k, \; \eeq
 $ \epsilon =$ diag$(1,-1,-1,-1)$ the signature diagonal matrix,  $\omega $ the orthogonally reduced   Levi-Civita connection,  and $\phi _l$ any matrix ``proportional''  to unity
\beq \phi _l = f_l \1 \; \; \; \;  \mbox{ with $ f_l$  real-valued function} \eeq
  \cite[264f.]{Fock:ZPh1929}. Fock thus arrrived at an explicit form of Weyl's infinitesimal spinor transformation $dE$, at least for the case of the (original) Dirac representation,
\[  dE \cdot \psi = \sum _l C_l dx^l \psi \, . \]

On that basis  Fock easily expressed  covariant differentiation of a spinor with respect to a vector direction of a the frame $\{e(\alpha )  \} $
\beq   D'_{\alpha } \psi = \frac{\partial}{\partial e(\alpha )} \psi - C_{\alpha } \psi \eeq
or a coordinate direction $x^j$
\beq D_j \psi  = \frac{\partial}{\partial x^j }\psi  - \tilde{C_j} \psi    \eeq
where   $ \tilde{C_j}$ are slightly different matrices calculated from the $C_j$'s. For Weyl, both versions of covariant differentiation could be derived from his ``covariant differential'' $\delta \psi $ of equation (8).

 \subsubsection*{An additional   $U(1)$-gauge}
Up to this point I omitted an important observation made by both authors, which led   back to Weyl's gauge idea. The ``lifting'' of the Levi-Civita connection to the spinor structure  was not uniquely determined, even if we neglect the double valuedness of the $SL(2,\C)$ covering of the Lorentz group. 

Fock's calculation of the the matrices  (equation (10)) showed that the compatibility condition determines the $C_l$ only up to addition of purely imaginary matrices 
$i f _l \1$.  Covariant differentiation of spinors (equations (12), (13)) is then affected by an additive term 
$- i f_{\alpha } \psi$ . In a kind of {\em d\'eja vu}  Fock realized that the additional term could be perceived as derived from a phase-gauge factor  of the $\psi$-field:
\begin{quote} 
The appearance of the Weylian differential form in the law of parallel displacement stands in close relation to the fact remarked by the author [Fock] and also by Weyl (\ldots) that the addition of a gradient to the 4-potential corresponds to a multiplication of the $\psi$-function by a factor of absolute value 1. \cite[266]{Fock:ZPh1929}
\end{quote} 
On that basis, Fock  formulated the Dirac equation for the general relativistic electron  by covariant derivation in his local spinor structure, including a Weylian $U(1)$-gauge  term as an integrated part of the covariant derivation (13) (ibid.)
\beq   F \psi = 0 \; \; \; \mbox{with \hspace{1mm}}
 F = i \h \sum_{j=0}^3 \gamma ^j D_j  +m c \gamma _4 \, .\eeq

Weyl discussed the question similarly, although slightly more general. He argued  that any semantically relevant information derived from a spinor field had to be invariant under $U(1)$-symmetries  of the spinor representation,  because  the $SO^+ (1,3)$-covariants used to represent physical quantities were  given by  Hermitian forms   $<\psi , A \psi>  $   and thus  were invariant under multiplication by  a phase factor $e^{i \lambda}$ of $\psi$.
Therefore the spinor  connection (``the infinitesimal linear transformation $dE$ of the $\psi$'') is determined  by the ``infinitesimal rotations'' $\omega $ of the reduced Levi-Civita connection only up to ``a purely imaginary multiple $i \cdot df$ of the unit matrix''. In other words, with $dE$ 
\[ dE' = dE + i df \1 \]
is also compatible with the underlying metric of GRT. Weyl concluded:
\begin{quote}
For the unique determination of the covariant differential $\delta  \psi$
of $\psi $ such a $df$ for each line element $\vec{P P'} = (dx)$ starting from $P$ is needed. \cite[263]{Weyl:ZPh1929}
\end{quote}
The selection among the spinor connections compatible with the Levi-Civita connection  could justly  be considered as a ``gauge'', in   strong analogy to the length gauge of 1918. 
 Morover Weyl used, just like Fock,  the possibility to express the Dirac equation of the electron in an electromagnetic field by means  of covariant differentiation of spinors {\em including} a $U(1)$-gauge potential (``such a $df$'').

 For action functions applying to spinor fields he felt it legitimate to postulate: 
\begin{quote}
If one (\ldots) substitutes
\[ 
\psi \; \mbox{by \hspace{2mm}} e^{i \lambda } \cdot \psi  \mbox{\hspace{8mm}}
f_p \;  \mbox{by \hspace{2mm}} f_p - \frac{\partial \lambda }{\partial x_p} \]
with $\lambda $ an arbitrary function of the position, gauge invariance necessarily holds, in the sense that the action principle remains invariant. \cite[263]{Weyl:ZPh1929}
\end{quote}

From the point of view of infinitesimal symmetries,
the new gauge structure resembled in certain features Weyl's study  of the {\em Raumproblem} early in the 1920s.  In the analysis of the space problem he had characterized   ``congruences''    by a subgroup $G$ of $SL(n,\R)$, contained in  a larger group $H$ of ``similarities'', in which  $G$ was  normal  (in fact,  $H$ was the normalizer of $G$ in $GL(n,\R)$). One  of his  postulates was a uniqueness condition for an affine connection equivalent (in a certain sense) to a given linear connection in the larger group. In 1929  he again dealt with a pair of groups, now given by physical considerations, the  smaller one being the Lorentz group or its universal convering, $G = SL(2,\C)$, and the larger one was
  $H = SL(2,\C) \times U(1)$ in which $G$  was normal by construction. Again a uniqueness condition for a connection, compatible to another given one,  played a crucial role for the analysis. 
The uniqueness condition was now formulated ``bottom up'', i.e. from a given (Levi-Civita) connection in the smaller group  to the larger one, and uniqueness of the (spinor) connection with respect to the larger group was achieved only by adding a connection in the quotient group $U(1)$ (respectively bundle, from the later point of view). In this sense there was a structural  analogy considering group extensions for infinitesimal symmetries, although the methodology had changed considerably. In 1929 Weyl  no longer tried to found his approach on a  priori principles, but rather analyzed 
  symbolic forms worked out (``constructed'') by  mathematical physicists in close communication with experimental knowledge of the rising quantum physics.

Weyl  discussed how one could  arrive at  physical consequences from his approach. It would lead us too far  to follow this line here.\footnote{Cf. \cite{Straumann:DMV}.}
I just want to mention  that Weyl drew impressive consequences  from the postulate of invariance of the action integral under infinitesimal symmetries of different kinds:
\begin{itemize}
\item[---] infinitesimal rotations of the frames leads to symmetry of the energy-momentum tensor,
\item[---] infinitesimal coordinate translations  leads to ``quasi''-conservation of energy and momentum and  in the case of special relativity by integration to invariance of rotational momentum   
\cite[256ff.]{Weyl:ZPh1929},\footnote{Weyl spoke of ``quasi-conservation'' of  energy-momentum  $t^q_p$,  because of a  second term in the differential equation derived from invariance under infinitesimal translations:
\[ \frac{\partial t^q_p}{\partial x_q} + \frac{\partial e^q (\alpha )}{\partial x_p  }t_q(\alpha )=0 \] Literal  conservation of energy and momentum holds  only if the respective terms of the gravitational fields are added or,  in special relativity, after specialization of the ONF's \cite[257f.]{Weyl:ZPh1929}.}
\item[---] infinitesimal $U(1)$ gauge transformations  leads to conservation of charge (ibid., 264f.).
\end{itemize}

He hoped, morover, that his general relativistic approach to the Dirac equation, together with the separation of the spinor fields into components of irreducible representations $\rho $ and $\rho^+ $ might lead to a solution of the problem of {\em negative energies} in the original Dirac equation.
In late 1929  Dirac proposed a solution to this problem  by some  imaginative ad-hoc arguments  postulating  the existence of positive electrons (positrons) appearing as constitutive parts of  the solution of the original Dirac equation with non-vanishing mass term, and  surprising  ``fluctuations'' between positive and negative charge contributions  to it. It turned out that neither the   positive charge contributions could  be separated nor the resulting  ``fluctuations'' 
eliminated from the solution \cite[90ff.]{Kragh:Dirac}.

Weyl, for his part,  attempted  for a short while in 1929 to avoid  such  fluctuations by the proposal  to study solutions of a modified  Dirac equation in the  irreducible components of the representation $\rho$ and $\rho^+$  separately (Weyl spinors). He remarked, however, that in this equation no mass term could be included   without losing gauge invariance \cite[242]{Weyl:NAS1929}. As a research strategy to overcome the problem he proposed  to neglect at first,  on the level of the spinor equation,   the mass of the electron and to reconstruct it,  in a second step of theory development, as an integral invariant that couples to  gravitation. 
 \begin{quote}
Be bold enough to leave the term involving mass entirely out of the field equations. But the integral of the total energy density over space yields an invariant, and  at the same time constant, mass; {\em require of it that its value be an absolute constant of nature m} which cannot vary in value from case to case. This introduction of mass is born of the idea that the inertia of matter is due to its energy content. \cite[243]{Weyl:NAS1929}
\end{quote} 
Such an approach made sense only in a joint theory of  gravitation, quantum physics (in the sense of the modified Dirac equation) and electromagnetism. In his attempt for an integrated theory Weyl now pursued the concrete goal  to contribute to the solution of  the mass problem of the electron.

The  proposal to start from a ``massless'' electron was rejected by physicists immediately. In the  postscript to his article for the {\em Physikalische Zeitschrift} Fock  argued strikingly (and presumably also convincingly for Weyl)\footnote{In the 2nd edition for {\em GQM} Weyl no longer insisted on his 1929 proposal and supported  Dirac's strategy to deal with  the problem \cite[2nd. edition, 230, 233]{Weyl:GQM}.}
that the current of the Weyl-spinor field was lying on the light-cone. Thus there remained  no realistic  hope for a solution of  the electron's mass problem along the line indicated by Weyl \cite[276f.]{Fock:ZPh1929}. 
Similarly  Pauli  rejected Weyl's proposal to circumvent the mass problem for the electron, although from a conceptual point of view  he found the new integration of the gauge idea into  quantum physics most convincing. He contributed essentially to its dissemination and survival in  the physics community. Moreover he revived Weyl spinors in 1956 when he looked for an adequate mathematical representation of his newest hypothetical entity, the neutrino.
This is a different and historically complicated story which cannot be dealt with here.\footnote{See \cite[313ff.]{Pais:Inward}, \cite{Straumann:DMV}.}

Weyl indicated that field quantization was  another problem that   had to be solved before one might hope for an answer to the questions raised:
\begin{quote}
Another difficulty which stands in the way of a comparison with experience is that the field equations must first be quantized before they can be applied as a basis for the statistics of quantum transitions. But our theory is also hopeful in this respect inasmuch as the anti-symmetric Fermi statistics of the electrons, corresponding to the Pauli exclusion principle, here necessarily leads to the symmetric Bose-Einstein statistics of photons. \cite[244]{Weyl:NAS1929}
\end{quote}
Weyl  could probably not surmise which tremendous difficulties had to be surmounted on the path  indicated here. When he reworked {\em GQM} for the second edition he knew already  more about the nature of problems arising from the infinities of  field quantization. He made some striking observations with respect to symmetries in quantum electrodynamics,  but  did not  contribute to its further development in the later 1930s and 40s.\footnote{For Weyl's contribution to the symmetries in early quantum electrodynamics, see \cite[287ff.]{Coleman/Korte:DMV}; for the history of quantum electrodynamics \cite{Schweber:QED}.}  

\subsubsection*{Geometry and physics: interpretations and perspectives}
As we have seen, Weyl's and Fock's 1929 work contained a strong common mathematical core. They both established local spinor structures on  Lorentz manifolds  with an additional internal $U(1)$ symmetry  and proposed to use a connection in this structure, determined by or determining  gravitation and electromagnetism and governing the motion of the spinor field. But they had strong differences with respect to  the question of how geometry and  physics could or  should   be related. 

Fock   proclaimed that  his goal was ``the geometrization of Dirac's electron theory and its subsumption (Einordnung) in general relativity''  \cite[275]{Fock:ZPh1929}. This  was   a  conceptual-methodological task, rather  than one  of concrete physical theory building. He hoped, however,  that
his investigation   might  ``contribute to the  solution of the problems'' in Dirac's theory, referring apparently to the paradox of negative energies and positive probability  of fluctuations between negative and positive energies, respectively charges. He thus expected that his geometrization of the  Dirac operator might lead,   in the long run,   to  progress of a physical theory in a more technical sense. 
Fock's main hope was, however, to contribute to what he (and Ivanenko) thought to be a challenging goal of contemporary physics, the development of a common conceptual structure for relativity and quantum physics.

  V. Fock had learned relativity from A. Friedmann and participated prominently in the development of relativity theory in Russia.\footnote{\cite[286ff.]{Gorelik/Vizgin}.} In the later 1920s he maintained  close contact to a group of young  physicists in Leningrad around L. Landau, G. Gamow, and M. Bronstein,  to which his early 1929 coauthor D. Ivanenko belonged.  The young physicists  enthusiastically supported the  cultural awakening in the early Soviet Union and wanted to contribute to it through  their work in relativity and  quantum physics.\footnote{\cite[20ff.]{Frenkel/Gorelik:Bronstein}.}  This was apparently part of the background for Fock's and Ivanenko's premature  claim to have found a path  towards {\em quantum geometry}.

 In a  letter to  {\em Nature}, dated March 21, 1929,  they announced  their  first, still very sketchy  ideas on ``linear geometry''  as a contribution to this  challenging task.\footnote{With respect to their purely formal ``linear form with matrix coefficients'' $ds = \sum_k \gamma _k dx_k$ (see above) they proclaimed: ``This linear $ds$ is connected with  Dirac's wave equation in the same way as the Riemannian $ds^2$ with the relativistic wave equation of the older theory. \ldots This linear geometry seems to furnish a basis on which a uniform theory of gravitation, radiation, and quantum phenomena is to be constructed'' \cite{Fock/Ivanenko:Nature}. For more details they referred to their forthcoming paper  \cite{Fock/Ivanenko:ZPh} }
In the  {\em Comptes Rendus} note of May 22, 1929, they shifted attention in their  ``g\'eom\'etrie quantique lin\'eaire''  from the ``matrix valued linear metric'' to parallel displacements and covariant differentiation  in a local spinor structure. Once more, they claimed to have found a method  to reconcile quantum physics with geometry
\begin{quote}
Il importe de signaler un point qui distingue les id\'ees expos\'ees dans cette Note de celles d'Einstein et de Levi-Civit\`a: c'est l'intervention des matrices-op\'erateurs dans les \'equations pour les quantit\'es purement g\'eom\'etriques. Gr\^{a}ce \`a cela on peut bien s'imaginer un champ \'electromagn\'etique dans un espace euclidien, ce qui \'etait impossible dans les autres th\'eories. \cite[1472]{Fock/Ivanenko:CR}
\end{quote}

In his later contributions Fock was more cautious and weakened  the claim to the more moderate one of  having pursued  ``the geometrization of Dirac's theory of the electron and its subsumption under the general theory of relativity''  \cite[275]{Fock:ZPh1929}. 
He  admitted that the ``difficulties which are inherent in Dirac's theory''  had not yet been touched,  but  added:
\begin{quote}
Our investigations might perhaps contribute indirectly to the solution of these difficulties, by showing what the original unchanged Dirac theory can achieve. (ibid.)
\end{quote}
The   reference to the ``original unchanged Dirac theory'' was probably formulated after Fock got to know  Weyl's proposal and indicated a  disassociation from the latter, the reasons of which were explained  in the postscript.  Fock thus proclaimed that  the geometrization  of the Dirac equation by the spinor structure  with connections and covariant derivation  was  an important methodological achievement in itself.

On this point Weyl did not agree at all. He had lost confidence in the geometrical unification programs  which he himself had contributed so effectively by  his gauge unification in 1918.   About the end of the 1920s he no longer expected  any deeper understanding of physical reality by the still blossoming geometrical unification programs.\footnote{On the ``diversity'' of these programs see \cite{Goldstein/Ritter:UFT}.} He criticized, in particular, Einstein's latest attempt at  unification by an additional structure of distant parallelism as a turn towards a physically unmotivated ``artificial geometry''  \cite[219 quoted above]{Weyl:NAS1929}. In his later 1929 paper for  {\em Physikalische Zeitschrift} he argued in more  detail:
\begin{quote}
I am unable to believe in distant parallelism for several reasons. Firstly, a priori,  my mathematical sense (mathematisches Gef\"uhl)  opposes against accepting such an artificial geometry; for me, it is  difficult  to conceive of a  power which  would make  the local systems of axes, in their twisted position in the different world-points, freeze together in rigid affiliation. Moreover, two important physical reasons have to be added.  \ldots \cite[246]{Weyl:ZPh1929}
\end{quote}
As ``first physical reason'', Weyl mentioned  his gauge theory of electromagnetism. He argued that only the point-dependence of the ONF's gave rise to a variable phase factor $e ^{i \lambda }$ and thus the new principle of gauge invariance. The ``second physical reason'' was, to Weyl, the possibility to derive symmetry of the energy-momentum tensor and the invariance of rotational momentum in special relativity from infinitesimal rotations of the ONF's or of infinitesimal translations of coordinates (see above). Thus Weyl's ``physical reasons'' consisted essentially of  methodological arguments for the superiority of invariance properties in  an infinitesimal symmetry approach, close to those  which about three decades later became central in the rise to prominence  of more general ``gauge'' theories.\footnote{Cf. \cite{Morrison:Symmetries}.}

The 1930  Rouse Ball lecture at  Cambridge university gave Weyl the opportunity  to explain his view of the unification programs to a wider scientific audience. He still considered the attempts  ``to geometrize the whole of physics'', undertaken after Einstein had   so  successfully  geometrized  gravitation,  very comprehensible at its time \cite[338]{Weyl:Rouse_Ball}. He explained  his own theory of 1918 and summarized its critical reception by physicists. He  reviewed Eddington's approach to unification by affine connections and Einstein's later suppport for that subprogram, always in comparison with his own   ``metrical'' unification of 1918, and concluded that in hindsight one could see  that both theory types were   ``merely   geometrical dressings (geometrische Einkleidungen) rather than proper geometrical theories of electricity''. He ironically added that the struggle between the metrical and affine UFT's (i.e. Weyl 1918 versus Eddington/Einstein)  had lost  importance, as in 1930  it could no longer be the question   which of the theories would ``prevail in life'', but only ``whether the two twin brothers had  to be buried in the same grave or in two different graves'' (ibid., 343).   
He again made clear that he could not find any argument in favour of  Einstein's distant parallelism approach, nor could he  find good prospects for the Kaluza-Klein approach.\footnote{The revival of Kaluza-Klein type theories in the 1980s happened in a completely different content of theory development. In this conference, moreover, P. Cartier argued that there are reasons  which  might  lead to a renewed   interest in the original form of Weyl's {\em purely infinitesimal geometry}  --- again  in a modified physical interpretion and theory context. }
 Weyl even accused Einstein's new theory  of ``breaking with the infinitesimal point of view. (\ldots ) 
The  result is to give away nearly all which has been gained in the transition from special to general relativity. The loss is not compensated by any concrete gain'' \cite[343]{Weyl:Rouse_Ball}.

Weyl perceived a  nearly complete scientific devaluation of the UFT's of the  1920s,  resulting from developments in the second part of the decade:
\begin{quote}
In my opinion the whole situation has changed during the last 4 or 5 years  by the detection of the matter field. All these geometrical leaps (geometrische Luftspr\"unge) have been premature, we now return to the solid ground of physical facts. \cite[343]{Weyl:Rouse_Ball}
\end{quote}
He continued to sketch the theory of spinor fields,  their phase  gauge and its  inclusion into the framework of general relativity along the lines of the 1929 articles. Weyl emphasized  that,  in contrast to the principles on which the  classical UFT's had been built, 
the new principle of phase gauge ``has grown from experience and resumes a huge treasury of experimental facts from spectroscopy'' (ibid. 344). He still longed for 
safety,  just as much  as at the time after the First World War, when he designed his first gauge unification. Now he no longer expected to achieve it by geometric speculation,  but tried to anchor it in  more solid grounds: 
\begin{quote}
By the new gauge invariance the {\em electromagnetic field now becomes a necessary appendix of the matter field, as it had been attached to gravitation in the old theory. } \cite[345, emphasis in original]{Weyl:Rouse_Ball}
\end{quote} 
Weyl made it very clear to his readers that he had changed his perspective. He no longer saw a chance in  attempts to derive matter in highly speculative approaches from mathematical structures devised to geometrize force fields; he now set out to search  forms for the  mathematical represention of matter, which gave expression to the enduring  traces in the ``huge treasury'' of experimental knowledge. For him, this was reason enough to prefer  the view that the electrical field ''follows the ship of matter as a wake, rather than gravitation'' (ibid.).

In short, {\em Weyl had turned from his idealist approach to matter}, pursued at the turn to the 1920s,   {\em to a symbolic realist} one at the end of the decade. This change of perspective  had consequences for his views on geometrization. With reference to Fock's interpretation of the role of geometry in the general relativistic Dirac equation Weyl continued:
\begin{quote}
Mr. Fock calls the derivation of the new gauge invariance from general relativity, which he arrived at nearly simultaneously with me, a geometrization of  Dirac's theory of the electron. In this respect I cannot agree with him. My impression is that we have abandoned geometrization  by linking electricity to matter rather than to gravitation. I fear that the geometrizing tendency,  which  seized gravitation  in full right and supported by the most intuitive arguments, was misled when it was extended to other physical entities. \cite[345]{Weyl:Rouse_Ball} 
\end{quote}
Weyl did not, on the other hand, completely negate any {\em  possibility } to find a   geometrical quantum theory.  He only warned that, if one wanted to continue with the geometrizing tendency,  one had to invent a ``natural geometry''   leading to a spinor type field $\psi$  for the characterization of its structure, in addition to the ONF.  Whereas Fock claimed to have achieved  this already, Weyl remained agnostic:
\begin{quote}
One had to set out in search of a geometrization of the matter field; if one succeeds here,  the electromagnetic field  is added as  a premium to the bargain. I have no idea what kind of geometry this might be. (ibid.)
\end{quote}
From the perspective of late 20th century developments in differential geometry and the tremendous role of gauge field theories, Weyl's evaluation   is highly surprising and even seems   paradoxical:  Why  did he  not percieve his own and Focks's invention of local spinor structures with additional $U(1)$-gauge as a sufficiently rich extension of geometry  to deal with matter structures?\footnote{I thank Jim Ritter  who indicated this point to me  and insisted on a closer historical perspective.}

Our own perspective has been shaped by 
 the  development of differential geometry and topology in the second half of the last century, which was   deeply influenced by Elie Cartan's work, the work of his students and other researchers. In the late 1950s and  1960s  bundle structures  with their inbuilt transformation behaviour have become  central  concepts in geometry and topology.   In this sense, Weyl's first desideratum  of a ``natural geometry''  which  includes  spinor type field in its core structure seems to be satisfied, and it  becomes difficult to grasp why Weyl, unlike Fock, did {\em not} accept their common contribution as a valuable step in this direction. 

We  may assume that Weyl over-emphasized his   scepticism with respect to geometrization of physics at the turn to the 1930s, because he still wanted to correct  his earlier  exuberance in this respect. Moreover he wanted to disassociate himself strongly  from the ``old'' unification programs  which where still alive in the latest attempts of Einstein,  or Kaluza and Klein, and wanted  to counteract them in the scientific discourse as clearly as possible.

For a proper historical understanding we have to take another  aspect into account.   
Weyl's attempts to integrate geometry with physics  had, from their very beginnings after the First World War, a strong intentional reference to the quantum stochastical aspects of matter as a a ``dynamical agens'',  even at a time when these were not understood at all. In the early 1920s Weyl had dared to    speculate in wide leaps about a possible relationship between the  intuitive, the mathematical and the physical understanding of the continuum, some inbuilt discrete ``free-choice'' structures and the end of classical determinism in  natural science.\footnote{Most prominent and controversially discussed  in this respect is \cite{Weyl:Kaus_Stat}.}  In 1925, in his manuscript for the Lobachevsky centenary volume (published only posthumously \cite{Weyl:Riemanns_Ideen}), Weyl indicated that the vagueness of physical determination of space-time localization has to be taken seriously for the basic  theoretical structure of geometry.  This vagueness ought to be considered  a principal feature for the mathematical characterization  of geometry and  to be dealt with,  in principle, in some stochastical approach informed by ``the actual state of physics'', i.e. quantum physics.  But then, so Weyl remarked, at a time when the ``new'' quantum mechanics was just being shaped,  the question, how  such a quantum stochastical foundation for geometry relates  to the   differentiable structure of classical geometry, turned into a  completely open problem. He ended the passage by the honest remark:
\begin{quote}
One has to admit that until now nearly nothing has been achieved for the  question what it means to apply differential calculus  to [physical] reality. \cite[12]{Weyl:Riemanns_Ideen}
\end{quote}
With such questions Weyl was not {\em completely} alone. But they were  far from what most physicists or mathematicians considered useful at the time, or even later in the 1930, when Fock's  young colleague M. Bronstein explored the questions of a necessary revision of 
time-space concepts from the point of view of quantum physics \cite[83ff.]{Frenkel/Gorelik:Bronstein}. Fock's    hope of 1929     to leave classical geometry behind and to  turn towards  geometrical quantum structures was comparably   innocent.  With  such a  point of view  he was  content with an  extension of differential geometry which would appear, at most, as a semi-classical enrichment.

 In his  1930 talk at Cambridge (and its later publication) Weyl expressed  clearly  that  from a proper {\em geometry of matter} he expected a deep break with the  classical tendency of geometrization prevailing in the UFT's.  He was less clear, to say the least, what should be substituted for it; but there were strong reasons for  such vagueness. 
His own approach to the mass problem of the electron had turned out to be unsatisfactory; Dirac's alternative appeared  more promising, but still had  a long way to go before  a technically valid solution of the quantization problem was in sight\footnote{See \cite{Schweber:QED}.} 
 ---  not to speak about the extensions of later quantum gauge field theories and the still unanswerable question of the mass spectrum of basic constituents of matter.
Therefore Weyl's remark  
``I have no idea what kind of geometry this might be'', was just as honest as
his comment  in 1925 that ``nearly nothing had been achieved'' for a semantically reliable relation of the differentiable structure of geometry to the ``actual state of physics''. 

Other contributions  to this conference  explore the much  broader and  deeper mathematical knowledge  at the turn to the 21st century. Notwithstanding  a whole range of new open questions and desiderata, including  the one for a historical evaluation of  recent developments, we now see  several candidate programs for a quantum geometry  aiming at (or preparing)  a unification of quantum field theories.\footnote{Two, at least, were presented to the conference (M. Atiyah and A. Connes), another one was planned (C. Rovelli).}
It is not yet clear, whether one of them (or perhaps  several) will ``prevail in life''.   Weyl's proposal to look for a ``geometry of matter'' informed by the treasury of experimental knowledge could still  be taken as an advice for a critical discourse  in and among the different research programs.\footnote{At the turn of the century we may add that, in addition  to  recent and coming  results in high-energy  spectroscopy,   geometrical aspects of low energy  EPR-type experiments  constitute a valuable novel  part of the ``treasury'' of experimental knowledge, which ought to be taken into account in a future ``geometry of matter''. }
Perhaps future  developments will  show  whether  Weyl's  guess  that  the geometrization of interaction and metrical fields is ``added as a bonus'' once a  proper geometry of matter has been achieved is just another  speculative dream. It still may turn out that it   indicates a hint  for an appropriate theory development.  

\vspace{30mm}

 \bibliographystyle{apsr}
  \bibliography{a_litfile}

\end{document}